\documentclass[aip,amsmath,amssymb,reprint]{revtex4-1}
\usepackage{xcolor}
\usepackage{graphicx}
\usepackage{dcolumn}
\usepackage{bm}

\usepackage[utf8]{inputenc}
\usepackage[T1]{fontenc}
\usepackage{mathptmx}
\usepackage{etoolbox}
\usepackage{adjustbox}
\usepackage{multirow}
\makeatletter
\begin{document}

\title[]{Structural and dynamical behavior of methane–water systems under nanoconfinement}

\newcommand{\UG}{División de Ciencias e Ingenierías, Campus León, Universidad de Guanajuato, Mexico}
\newcommand{\UVigo}{Departamento de Física Aplicada, Universidade de Vigo, E-36310 Vigo, Spain.}

\author{Jos\'e Torres-Arenas}
\affiliation{\UG}
\affiliation{\UVigo}

\author{\'Angel M. Fern\'andez-Fern\'andez}
\affiliation{\UVigo}
\author{Mart\'in P\'erez-Rodr\'iguez}
\affiliation{Instituto de Química Física Blas Cabrera, CSIC, E-28006 Madrid, Spain.}
\author{Manuel M. Piñeiro}
\affiliation{\UVigo}
\affiliation{Modelizaci\'on y Simulac\'on de Materiales Nanoestructurados, Univ. de Vigo, Unidad Asociada al CSIC, Spain.}

\date{\today}

\begin{abstract}
We investigate the structural and dynamical behavior of methane–water systems under nanoconfinement using molecular dynamics simulations across pore widths from 1 to 5 nm. Structural analysis reveals a strong and non-monotonic dependence on confinement: while tetrahedral ordering partially recovers as confinement is reduced, cubic-like order associated with clathrate precursors is maximized at intermediate pore sizes. Radial distribution functions show that three-dimensional correlations are suppressed under strong confinement, whereas lateral ordering persists, indicating a reduction in the effective dimensionality of structural organization.
Transport properties reflect the same structural competition. Parallel diffusion is non-monotonic with pore size, while perpendicular motion is subdiffusive due to confinement-induced trapping and heterogeneity. Methane exhibits stronger subdiffusion and remains dynamically coupled to the water matrix.
A characteristic confinement length scale emerges at which structural ordering, dynamical heterogeneity, and solvent–solute decoupling are simultaneously maximized. At strong confinement, three-dimensional correlations are suppressed, leading to dimensional reduction, frustrated ordering, and inhibited nucleation.
\end{abstract}

\maketitle 

\section*{Introduction}

Methane hydrate formation in confined environments has attracted considerable attention due to its relevance in natural and engineered porous systems.\cite{borchardt2018,both2021,casco2015} In marine sediments, methane hydrates often develop within nanoporous media, where confinement and fluid–surface interactions significantly modify the structure and dynamics of water and methane.\cite{cuadrado2018,mi2022,yan2016} At the nanoscale, solid interfaces induce density oscillations, molecular layering, and strong anisotropy in transport, which can alter hydrate stability and growth mechanisms relative to bulk conditions. A molecular-level understanding of these effects is therefore essential for describing hydrate behavior in realistic porous environments.\cite{Bagherzadeh20123188,Zheng2022124718,Zhang2022, Zhang2023,Fan2024,Li2026}\\

Molecular dynamics simulations have become a key tool to investigate these systems, providing direct access to microscopic structure and dynamics. Previous studies have shown that confinement can modify hydrogen-bond networks, methane solubility, hydrate stability conditions, and spatial heterogeneity of confined fluids\cite{Walsh20091095,Sarupria20122942,Khurana201711176,Chen2021,Zhang2025}. Particular attention has been given to interfacial water layers, which often exhibit distinct structural and dynamical properties compared to the pore interior.\cite{Koga2000564,Shepherd201212172,Conde2008} Confinement also leads to pronounced dynamical anisotropy, especially in slit-pore geometries, where transport parallel and perpendicular to the walls becomes strongly decoupled.\cite{Mosaddeghi2012,Spera2020}. Despite these advances, the relationship between interfacial layering, structural ordering, and anisotropic transport, as well as its dimensional character under strong confinement, remains incompletely understood.\\

In previous works\cite{Fernandez2022JML, Fernandez2024JCP}, we studied methane hydrate growth in atomistic silica nanopores, identifying interfacial layering, structural defects, and confinement-induced distortions during crystallization. More recently, we examined the emergence and suppression of hydrate-like ordering under nanoconfinement, revealing regimes where three-dimensional structural organization is destabilized.\\

A related Communication has recently reported a complementary analysis of methane hydrate nucleation under similar confinement conditions, focusing on structural signatures of hydrate-like ordering and the dimensional reduction of molecular correlations.\cite{torresarenas2026} The present work builds on that study by extending the analysis to include interfacial structure and local ordering in greater detail, as well as the anisotropic diffusion of both species and the associated dynamical decoupling, which were not addressed in the shorter report.\\

Here, molecular dynamics simulations of methane–water mixtures were performed in planar silica nanopores with widths ranging from 1 to 5 nm and temperatures between 250 and 295 K. The discussion focuses on representative pore sizes (1, 3, and 5 nm) and selected temperatures that capture the transition between structurally ordered and disordered regimes. This choice allows us to isolate the main physical trends, including the suppression of three-dimensional ordering under strong confinement and the emergence of localized structural organization at intermediate pore sizes.\\

To characterize these effects, we combine density profiles, three-dimensional and in-plane radial distribution functions, local order parameters $F_3$ \cite{BAEZ1994177} and $F_4$ \cite{Rodger1996326}, and diffusion coefficients parallel and perpendicular to the confining walls.\\

The projected radial distribution functions provide direct insight into the dimensionality of structural correlations, distinguishing between three-dimensional ordering and confinement-induced quasi-two-dimensional organization. We focus on the interplay between interfacial structuring, local order, and anisotropic transport, and their dependence on pore size and temperature.\\

By combining structural and dynamical observables across a broad range of pore sizes and temperatures, this work seeks to identify the mechanisms by which nanoconfinement controls the emergence of local order, the dimensionality of molecular correlations, and the transport properties of methane–water mixtures. Establishing these relationships provides a microscopic framework for understanding the conditions under which confinement promotes or frustrates hydrate-like organization in porous environments.\\

\section*{System and Simulation Details}

This section describes the molecular models and simulation protocols employed to investigate methane–water mixtures confined in silica slit pores. Particular attention is given to the representation of the confining surfaces, the intermolecular interaction potentials, and the simulation conditions used to explore the effects of pore size and temperature on the structural and dynamical behavior of the confined fluids.\\

A slit pore of $\alpha$-silica was modeled with an all-atom force field. In accordance with this model, the bonded interactions include 1-2 bonding, 1-3 bending vibration, and the 1-4 dihedral torsion interactions. The dispersive and electrostatic forces are defined with the Lennard-Jones and Coulomb terms, respectively. The bonded parameters involved in this framework were published by Azenha \emph{et al.}\cite{Azenha20115062}, while for non-bonded interactions, the parameters were taken from the work of Smith \emph{et al.} \cite{Smith200420340}. The unit cell of $\alpha$-silica was obtained from the American Mineralogist Crystal Structure Database\cite{Downs2003247} and It was replicated 8 $\times$ 8 $\times$ 4 obtaining a dimension of 4.022 $\times$ 3.483 $\times$ 2.265 nm. As Periodic Boundary Conditions (PBC) are applied in three dimensions, the crystal structure is continued along X and Y dimensions, but is truncated in Z axis. This means, that silicon and oxygen atoms remain unsaturated in the surface area. Therefore, unsaturated silicon atoms are removed and terminal oxygen atoms are saturated with hydrogen. In this way, the neutrality of the system and a hydrophilic surface are obtained. This block of silica contains 1536 oxygen, 704 silicon and 256 hydrogen atoms. A more detailed description of the $\alpha$-silica structure and how to obtain this hydrophilic surface can be found in our previous works\cite{Fernandez2022JML,Fernandez2024JCP}. The OPLS-AA force field\cite{MacKerellJr.19983586} was used for modeling the methane molecule , and the well-known TIP4P/Ice\cite{Abascal2005Ice} for water. The use of those force fields led to good results with a reasonable computational cost.

The planar pores were opened enlarging the Z axis of the simulation box with the size required. In this way, systems featuring pore width of 1, 3 and 5 nm were obtained. This silica block is repositioned in Z axis leaving the empty cavity in the middle of the simulation box and enclosed by two hydrophilic surfaces. Each pore is filled with methane and water molecules. As in our previous works,  methane molecules are initially placed next to both silica walls and water molecules in the middle. The proportion between water and methane are shown in Table \ref{tab:pore-mol}. The different pore sizes maintain the same proportion of molecules.

\begin{table}[!h]
    \centering
    \begin{tabular}{ccccc}
        \shortstack{Pore\\Size} &  \shortstack{Lineal\\density (z)} & \multicolumn{3}{c}{Number of molecules}\\
         L$_z$ (nm) & d$_z$(molec/nm) & Water & Methane & Total\\
         \hline
         1 & 443 & 435  &   8 &  443\\
         2 & 443 & 831  &  56 &  887\\ 
         3 & 443 & 1210 & 120 & 1330\\
         4 & 443 & 1589 & 184 & 1773\\
         5 & 443 & 1969 & 248 & 2217\\
    \end{tabular}
    \caption{Number of water and methane molecules and in each pore and proportion of water molecules per nm length in Z direction.}
    \label{tab:pore-mol}
\end{table}

The leap-frog algorithm \cite{Frenkel2001} is used to solve Newton’s equations with a time step of 2 fs. LINCS algorithm\cite{Hess19971463} was applied for constraining all bonds. A cut-off radius of 1.5 nm for the dispersive and the real part of Coulomb forces was used. Far away of this distance the electrostatic interactions were addressed using Ewald sums method\cite{EssmannJCP1995}. The simulations were launched in the NVT ensemble where particles (N), Volume (V) and temperature (T) are maintained constant. As this is not the natural ensemble for molecular dynamics a thermostat is necessary to control the temperature. In our simulations, the Nosé-Hoover thermostat\cite{NoseJCP1984,HooverPRA1985} was chosen for keeping the temperature constant with a couple time of 0.2 ps. 

The energy of the simulation box was minimized using the Steep descendent algorithm. Then, a little equilibration step of 100 ps was carried out in the NVT ensemble at the temperature desired. Finally, the simulations were launched at different temperatures depending on the pore size. The thermodynamic conditions of all simulations fall inside the methane hydrate stability zone for the force field employed and they all were run with GROMACS software package 2023 version\cite{Abraham201519,Pall2020}.

\section*{Results and Discussion}

\subsection*{Density profiles and interfacial layering}
We begin by analyzing the spatial organization of the confined fluids through their density profiles across the pore. We examine how confinement and temperature affect interfacial layering and molecular structuring for both water and methane, focusing on representative pore sizes (1, 3, and 5 nm) that capture the transition from strong confinement to more bulk-like conditions. For each species, density profiles are shown for three pore sizes in separate subpanels. In each case, two temperatures are selected to contrast regimes with and without pronounced three-dimensional ordering. The exception is the 1 nm pore, where no such ordering is observed at any temperature; here, the most widely separated temperatures are used instead.

\begin{figure}[h!]
\includegraphics[width=\columnwidth]{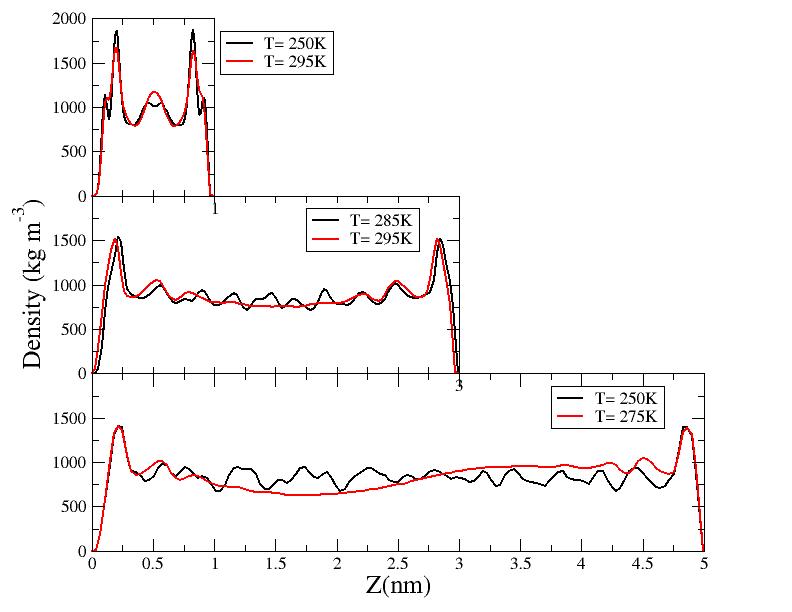}
\caption {Density profiles of water across the pore for representative pore sizes (1, 3, and 5 nm). Each panel shows the density variation along the confinement direction for two selected temperatures chosen to contrast regimes with and without pronounced structural ordering. In the 1 nm pore, where no three-dimensional ordering is observed at any temperature, the two most separated temperatures are shown. The profiles highlight the evolution of interfacial layering with pore size and temperature.}
\label{profileswater}
\end {figure}

\begin{figure}[h!]
\includegraphics[width=\columnwidth]{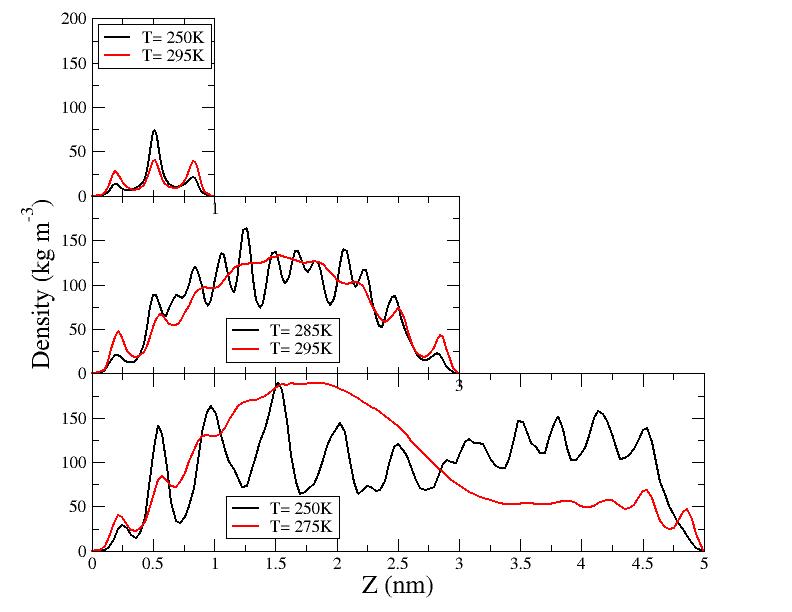}
\caption {Density profiles of methane across the pore for representative pore sizes (1, 3, and 5 nm). Each panel shows the density variation along the confinement direction for two selected temperatures chosen to contrast different structural regimes of the surrounding water. In the 1 nm pore, where no three-dimensional ordering is observed at any temperature, the two most separated temperatures are shown. The profiles illustrate how confinement-induced structuring of the host water affects methane interfacial localization and layering.}
\label{profilesmethane}
\end {figure}

Figure \ref{profileswater} shows the water density profiles across the pore for representative pore sizes of 1, 3, and 5 nm, highlighting the evolution of interfacial layering under confinement. In the 1 nm pore (top panel), the density displays two symmetric interfacial maxima near the walls and a weaker central peak, indicating a fully interface-dominated fluid with no bulk-like region. At intermediate confinement (3 nm, middle panel), a marked temperature dependence is observed: both temperatures exhibit pronounced interfacial peaks, but at 285 K a well-defined multilayered structure develops, whereas at 295 K the layering is reduced and progressively decays toward a shallow central minimum. For the 5 nm pore (bottom panel), interfacial peaks remain prominent, although the internal structuring becomes less regular. At 250 K, multiple oscillations persist across the pore, while at 275 K the profile becomes smoother, with damped layering connecting both interfaces.

Figure \ref{profilesmethane} presents the corresponding methane density profiles for the same pore sizes. In the 1 nm pore (top panel), the density is characterized by a dominant central maximum with weaker interfacial peaks and a slight asymmetry, indicating preferential occupation of the pore center with additional adsorption at the walls. At intermediate confinement (3 nm, middle panel), the profiles become strongly structured: at 285 K, pronounced multilayering extends from the interfaces toward the center, whereas at 295 K the structure is significantly smoother, with only weak residual oscillations and a broad central maximum. In the 5 nm pore (bottom panel), interfacial structuring persists but becomes less regular; at 250 K, several oscillations are still observed, while at 275 K the profile exhibits reduced layering and a weakly shifted maximum.

These results highlight the combined role of confinement and temperature in governing interfacial layering and spatial organization, with maximal structural ordering occurring at intermediate confinement and low temperature.

\subsection*{Local structure and orientational ordering}
To further characterize the structural organization of confined water, we now examine local molecular correlations and orientational ordering under confinement. In this subsection, water-water radial distribution functions together with the local order parameters $F_3$ and $F_4$  which quantify tetrahedral coordination and cubic-like local ordering, respectively, are analyzed in order to identify how confinement and temperature modify short-range structure, tetrahedral organization, and the degree of local ordering within the pore.\cite{steinhardt1983}\\

The $F_3$ and $F_4$ order parameters measures the spacial distribution of water atoms and two both allow to know how ordered are water molecules in our system. The $F_3$ order parameter evaluates the tetrahedral environment of water molecules by means the Equation \ref{eq:f3}

\begin{equation}\label{eq:f3}
    F_3=\langle(cos\theta_{icj}\mid cos\theta_{icj}\mid + cos^2\theta_t)^2\rangle
\end{equation}

$\theta_{icj}$ is the angle among the oxygens of three neighbor water molecules and $\theta_t$ is the tetrahedral angle. This expression averaged for the all oxygen atoms of all water molecules yields a value which is more approached to zero as the environment of the water molecule becomes more tetrahedral, and deviates from zero as the molecules become more disordered. So, $F_3$ allows to distinguish between liquid water and solid water, whether ice or hydrate. For this reason, the $F_4$ order parameter is highly encouraged to be computed since it can discern among liquid water, ice and hydrate. The Equation \ref{eq:f4} is used for obtained the $F_4$ value and its value is a function of the torsion angle between the outermost hydrogen atoms of two water molecules linked by a hydrogen bond.

\begin{equation}\label{eq:f4}
    F_4=\langle cos3\phi\rangle
\end{equation}

$\phi$ is the torsion angle. The averaged value of $F_4$ yields a value of -0.04 for liquid water, -0.4 for ice and 0.7 for hydrates.

\begin{figure}[h!]
\centering
\begin{minipage}[b]{0.52\textwidth}
         \includegraphics[width=\columnwidth]{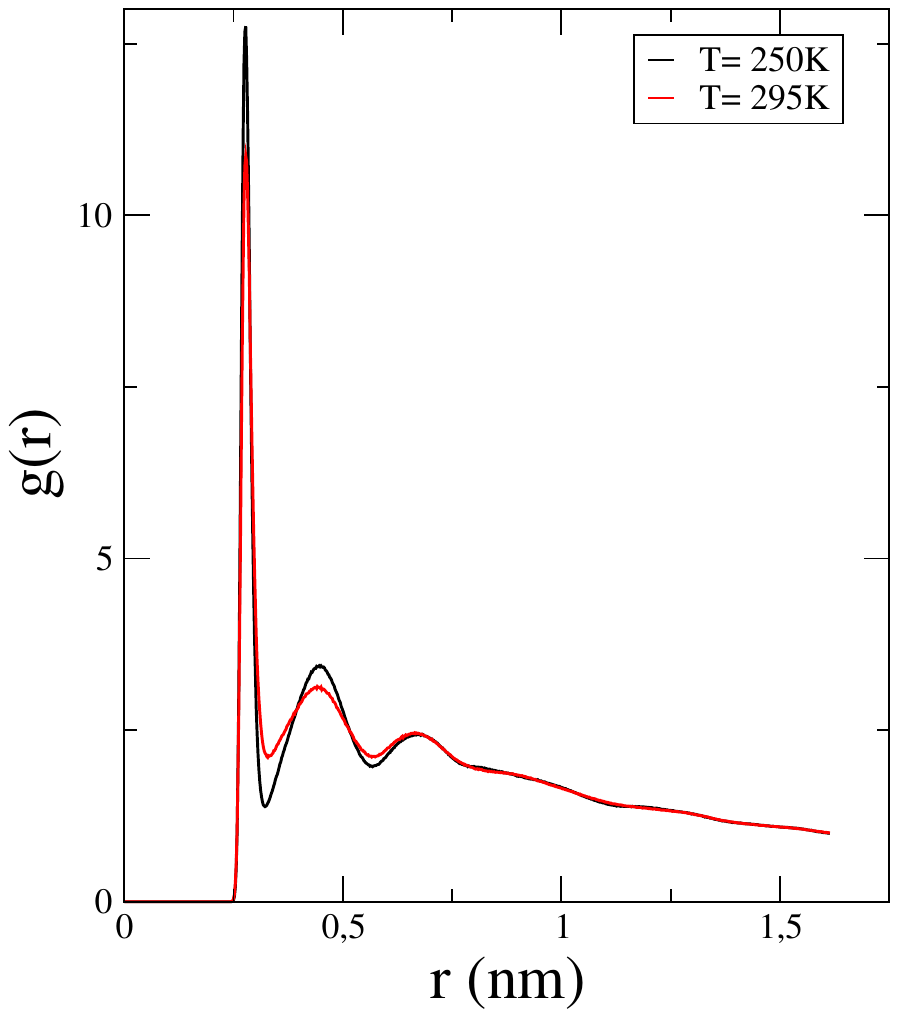}
\end{minipage}
\hfill
\begin{minipage}[b]{0.52\textwidth}
         \includegraphics[width=\columnwidth]{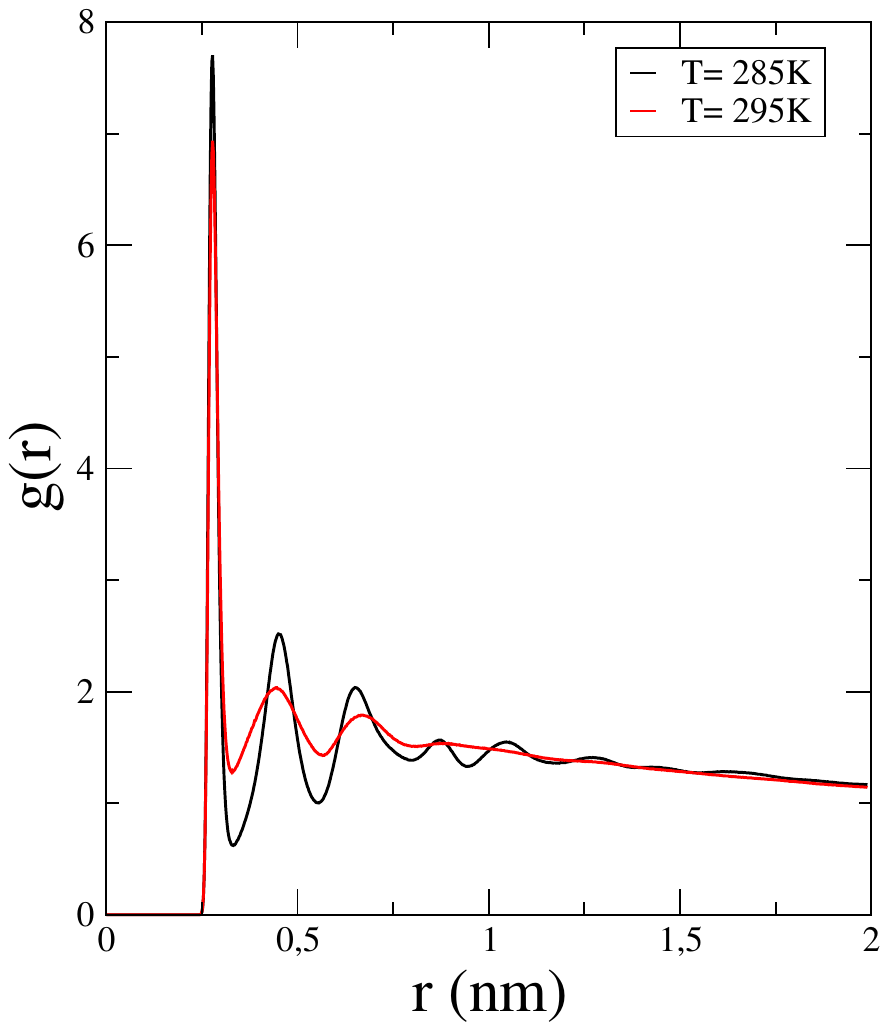}
\end{minipage}
\hfill
\begin{minipage}[b]{0.52\textwidth}
         \includegraphics[width=\columnwidth]{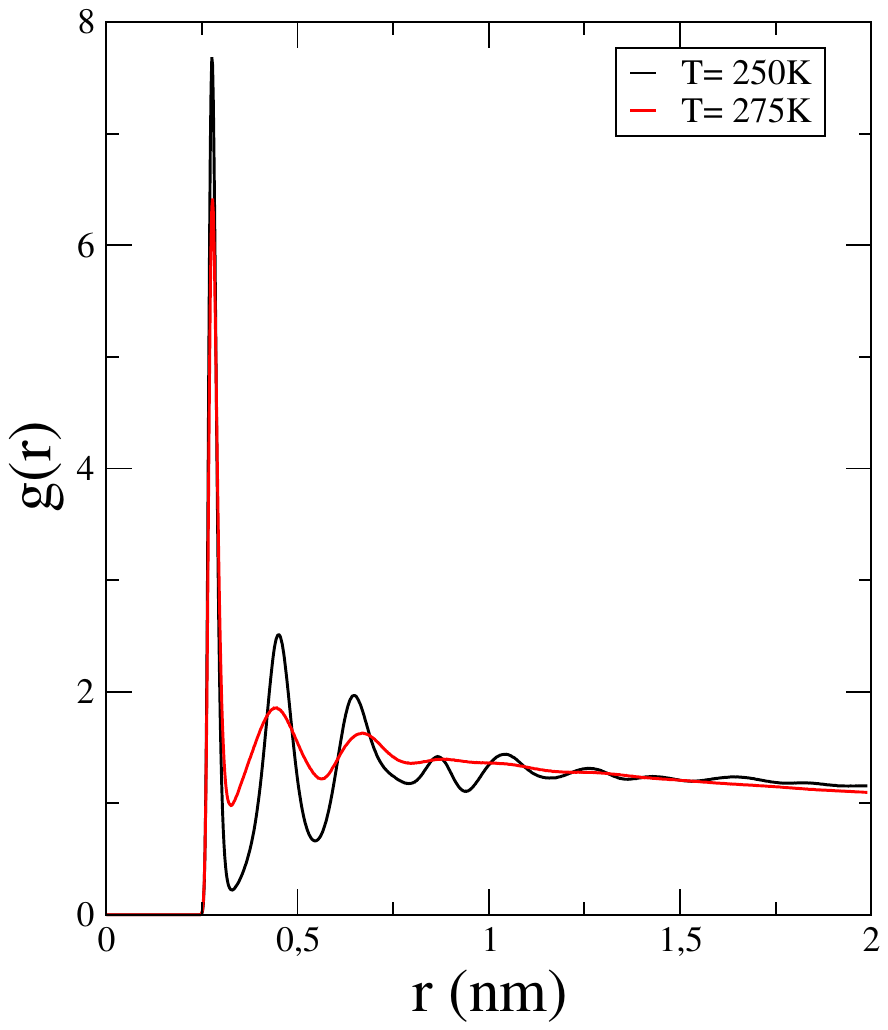}
\end{minipage}
\caption {Water--water radial distribution functions (RDFs) for representative pore sizes of 1 nm (top), 3 nm (middle), and 5 nm (bottom). Each panel compares two temperatures chosen to contrast structurally ordered and disordered regimes. For 1 nm, where no solid-like ordering is observed, the most extreme temperatures (250 and 295 K) are shown. The RDFs highlight the effect of confinement and temperature on local water structure.}
\label{rdfswater}
\end {figure}

\begin{figure}[h!]
\centering
\begin{minipage}[b]{0.52\textwidth}
         \includegraphics[width=\columnwidth]{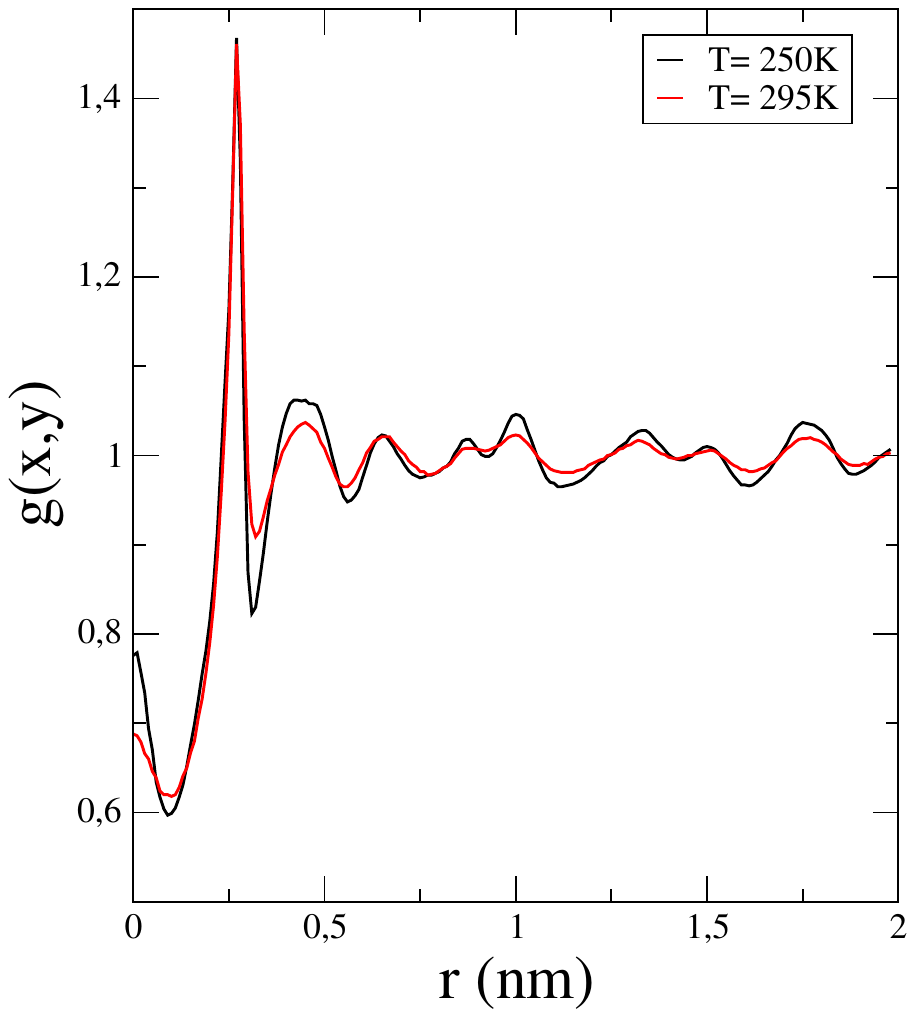}
\end{minipage}
\hfill
\begin{minipage}[b]{0.52\textwidth}
         \includegraphics[width=\columnwidth]{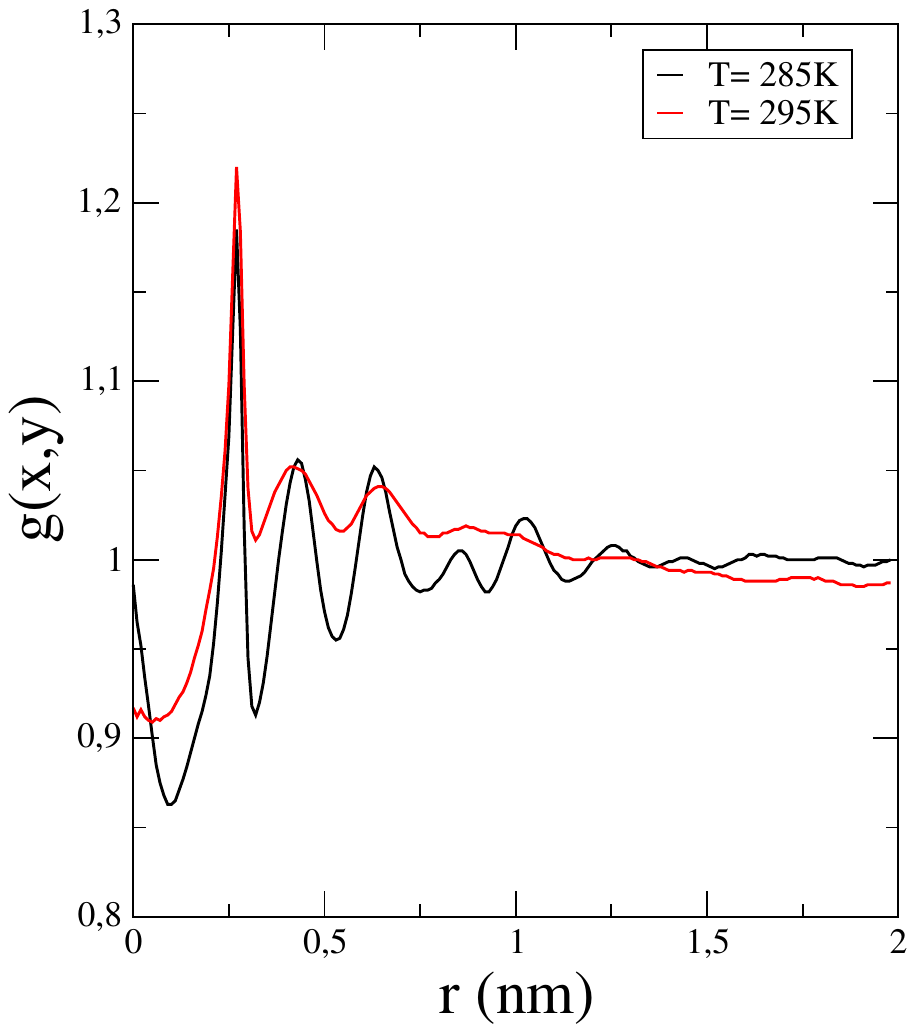}
\end{minipage}
\hfill
\begin{minipage}[b]{0.52\textwidth}
         \includegraphics[width=\columnwidth]{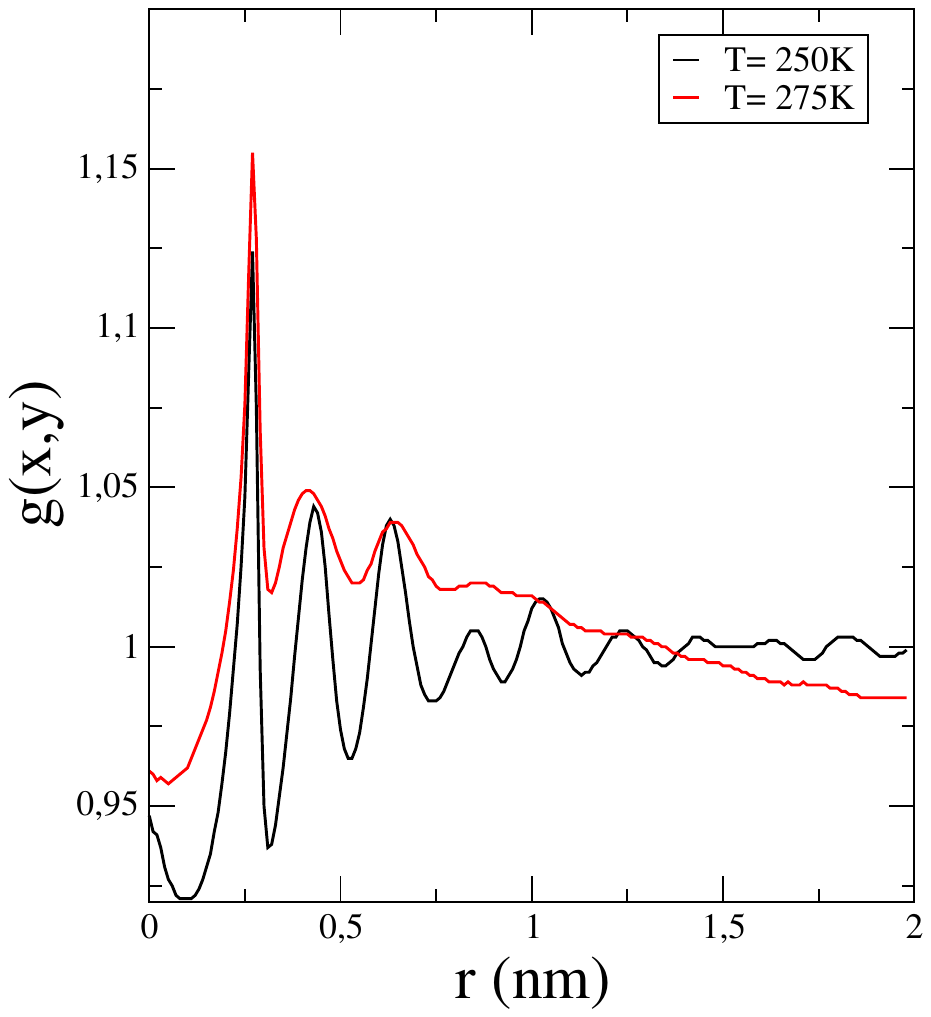}
\end{minipage}
\caption {Water-water radial distribution functions projected onto the $xy$ plane for pore sizes of 1 nm (top), 3 nm (middle), and 5 nm (bottom). Two temperatures are shown in each panel to contrast ordered and disordered regimes. The projected RDFs highlight the lateral structural correlations induced by confinement within the pore.}
\label{rdfsxy}
\end {figure}

Figure~\ref{rdfswater} shows the water–water radial distribution functions computed from molecular centers of mass for pore sizes of 1 nm (top), 3 nm (middle), and 5 nm (bottom), each case comparing a lower temperature where solid-like structuring is observed with a higher temperature where such ordering is absent. In the 1 nm pore, the RDFs display broadened and weakly defined peaks at both temperatures, with limited oscillatory structure beyond the first coordination shell. The similarity between the profiles indicates that lowering temperature does not significantly enhance structural correlations, reflecting the dominance of strong geometric confinement that suppresses extended hydrogen-bond networks and precludes three-dimensional organization.
At 3 nm, a clear temperature-dependent transition is observed: at lower temperature, the RDF exhibits sharper peaks and well-resolved coordination shells extending to larger distances, indicating enhanced short- and intermediate-range order consistent with solid-like structuring. At higher temperature, the peaks broaden and oscillations decay more rapidly, reflecting a predominantly liquid-like structure. A similar trend is observed at 5 nm, where low-temperature RDFs display well-defined peaks resembling bulk-like correlations, while at higher temperature the oscillatory structure becomes damped.
Additional insight is provided by the projected RDFs in the $xy$ plane (Fig.~\ref{rdfsxy}). In the 1 nm pore, pronounced oscillations persist at both temperatures, indicating well-defined in-plane ordering despite the absence of three-dimensional correlations. In contrast, for the 3 and 5 nm pores, the projected RDFs closely follow the behavior observed in the full three-dimensional analysis: strong oscillations at low temperature and smoother, liquid-like profiles at higher temperature, indicating that structural correlations retain a three-dimensional character.
These results show that confinement controls both the strength and dimensionality of structural correlations. While intermediate and weak confinement allow the development of three-dimensional order within a narrow temperature window, strong confinement suppresses bulk-like correlations and restricts ordering to in-plane structures, providing clear evidence of confinement-induced frustration of three-dimensional organization.

\begin{figure}[h!]
\centering
\begin{minipage}[b]{0.52\textwidth}
         \includegraphics[width=\columnwidth]{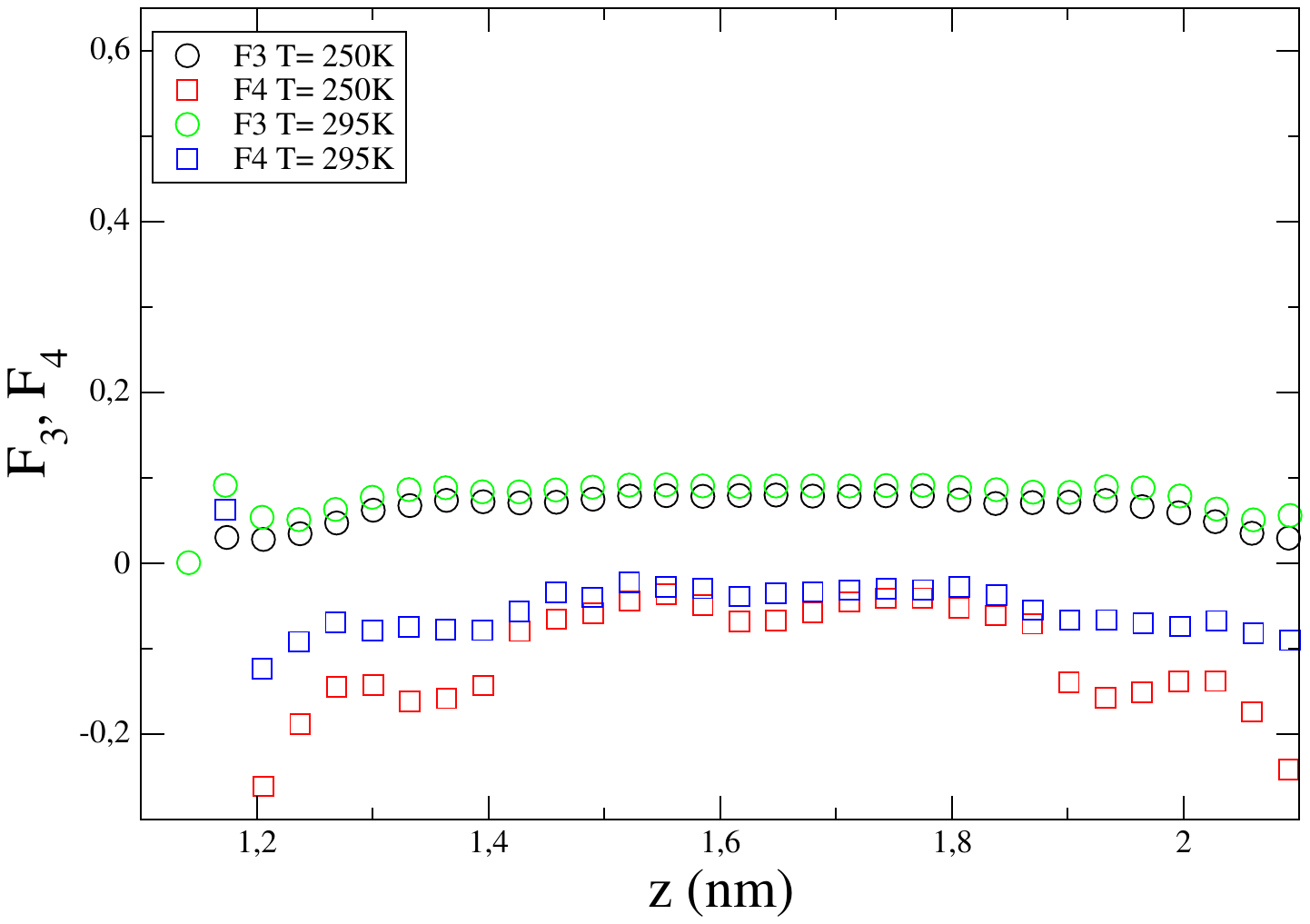}
\end{minipage}
\hfill
\begin{minipage}[b]{0.52\textwidth}
         \includegraphics[width=\columnwidth]{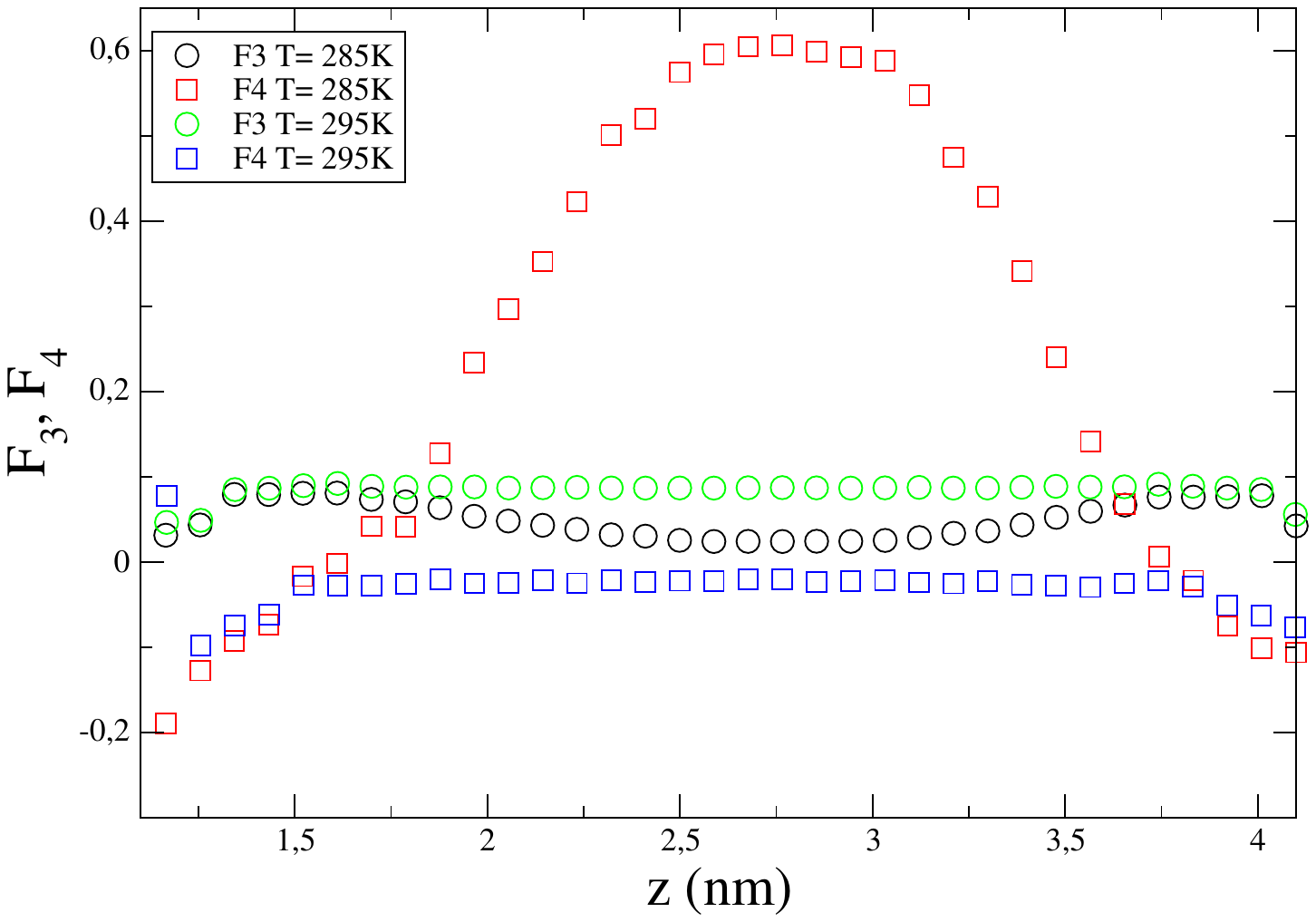}
\end{minipage}
\hfill
\begin{minipage}[b]{0.52\textwidth}
         \includegraphics[width=\columnwidth]{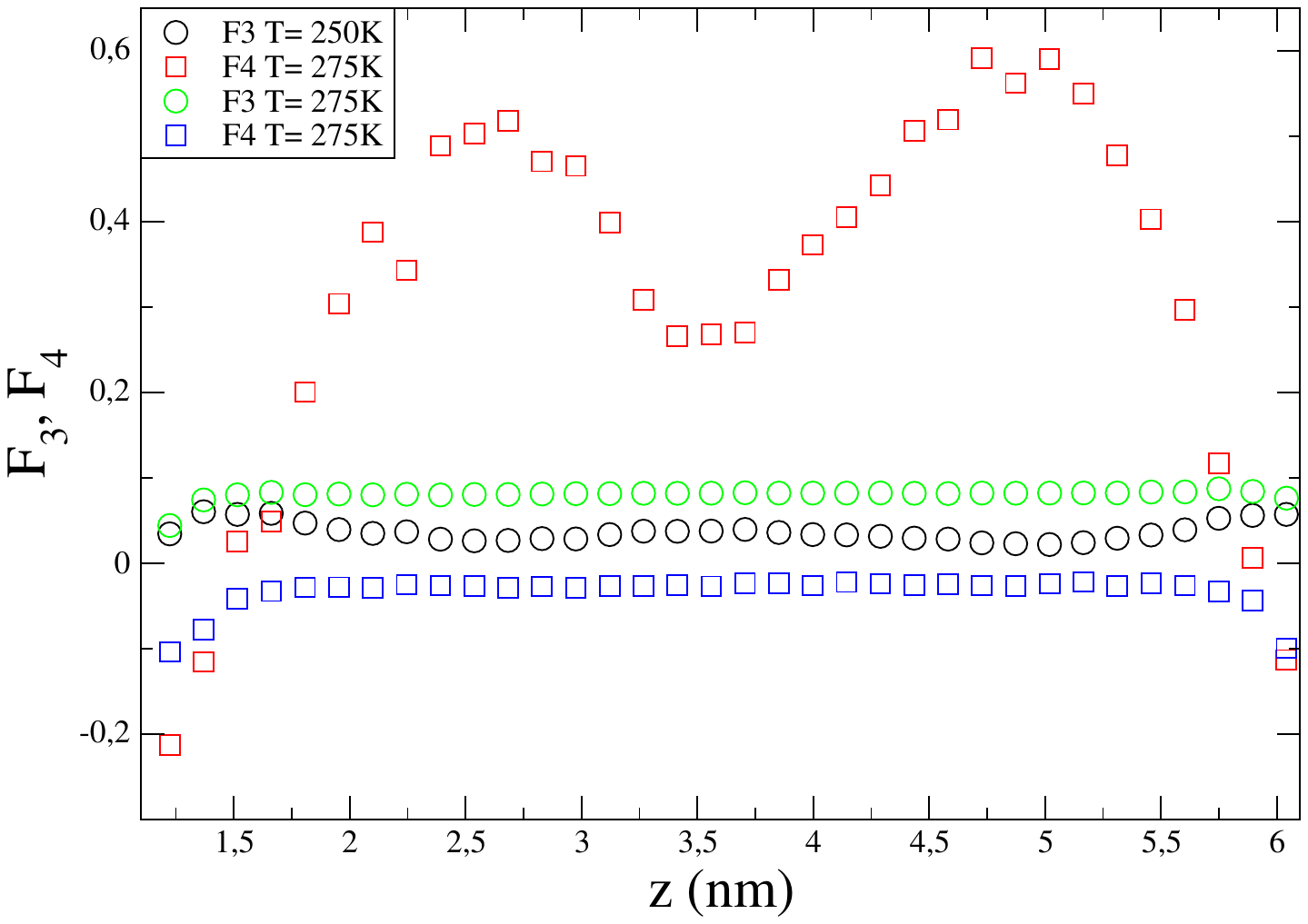}
\end{minipage}
\caption {Local structural order parameters $F_3$ and $F_4$ for water across the pore, shown for representative pore sizes of 1 nm (top), 3 nm (middle), and 5 nm (bottom). Each panel includes both order parameters for two selected temperatures. The comparison highlights the effect of confinement and temperature on local tetrahedral ordering and structural organization within the pore.}
\label{f3f4}
\end {figure}

Figure~\ref{f3f4} shows the spatial profiles of the local order parameters $F_3$ and $F_4$ for pore widths of 1 nm (top), 3 nm (middle), and 5 nm (bottom). In the 1 nm pore, both parameters remain low and nearly uniform across the pore, with weak temperature dependence, indicating strong disruption of tetrahedral coordination and absence of cubic-like ordering. Consistent with the RDF analysis, the lack of short- and intermediate-range correlations prevents the development of three-dimensional structural organization.
At 3 nm, both parameters develop a clear spatial dependence. The increase of $F_3$ toward the pore center indicates partial recovery of tetrahedral coordination away from the interfaces, while finite $F_4$ values at low temperature reveal the emergence of locally ordered environments with cubic-like character. These features are strongly reduced at higher temperature, indicating that such structures are spatially confined and thermally unstable.
For the 5 nm pore, the profiles approach a weak confinement regime while retaining interfacial modulation. At low temperature, elevated $F_3$ values indicate extended tetrahedral coordination, and enhanced $F_4$ suggests increased propensity for cubic-like local ordering. However, spatial variations persist, showing that these ordered regions remain fragmented and influenced by the interfaces. Increasing temperature reduces both parameters, reflecting progressive destabilization of local order.

Taken together, these results show that confinement governs the development of structural order in a non-monotonic manner. While tetrahedral coordination progressively recovers as confinement weakens, cubic-like ordering is maximized at intermediate pore sizes, reflecting a competition between confinement-induced structuring and geometric frustration that limits the formation of extended three-dimensional order.

\subsection*{Anisotropic diffusion coefficients}

Finally, in this section, we analyze the transport properties of confined water and methane under nanoconfinement, focusing on the anisotropic and non-classical dynamics induced by spatial restriction and temperature. In particular, we examine diffusion coefficients parallel and perpendicular to the confining surfaces, together with the associated anomalous diffusion exponents, in order to characterize deviations from standard Brownian motion. This framework allows us to relate the emergence of anisotropic transport to the structural heterogeneity and confinement-induced frustration discussed in the previous sections.

\begin{figure}[t]
\includegraphics[width=\columnwidth]{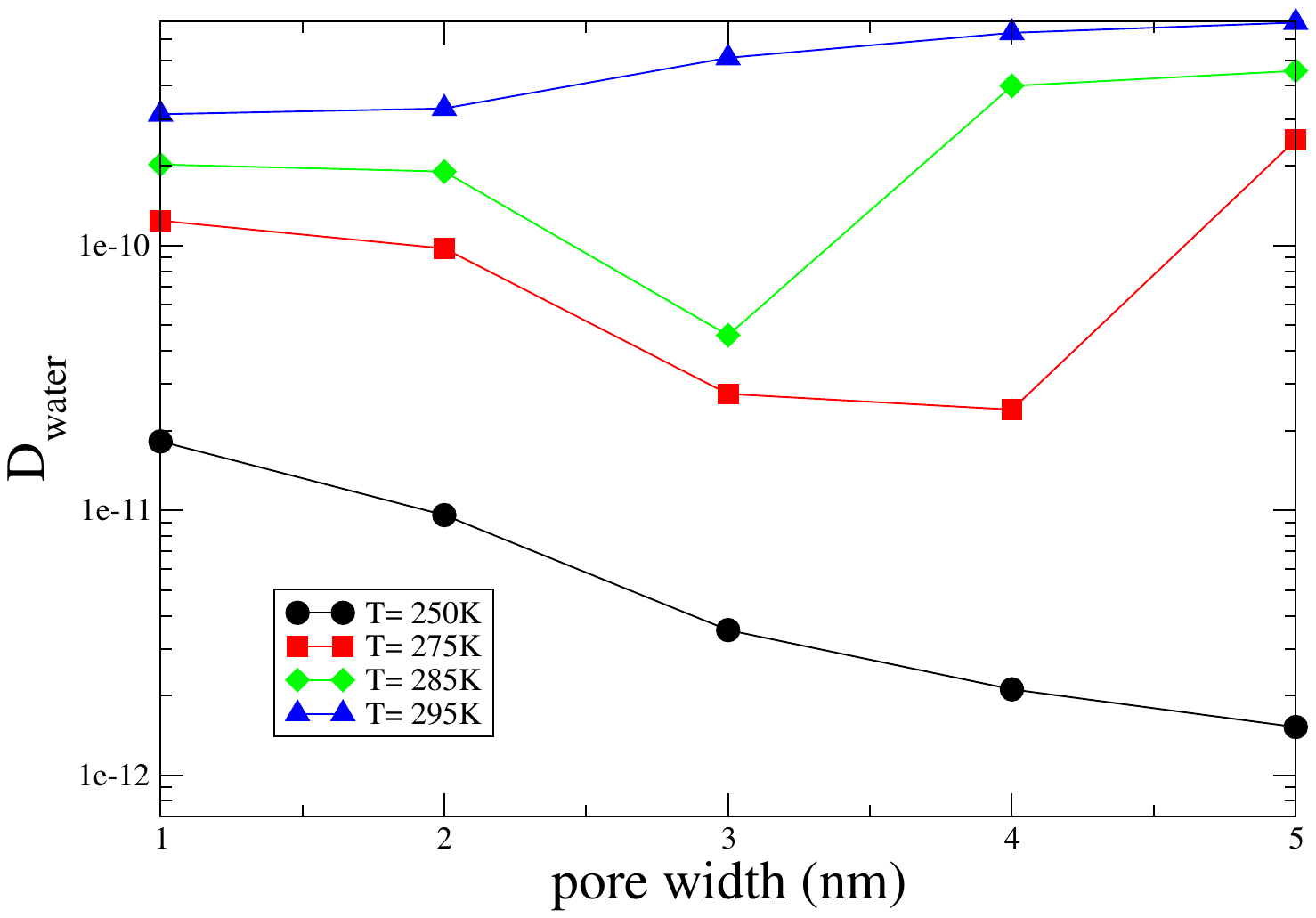}
\caption{Parallel diffusion coefficients of confined water as a function of pore size for temperatures $T=250$, $275$, $285$, and $295$ K. The diffusion coefficients are shown on a logarithmic scale. From bottom to top, the datasets correspond to increasing temperature. Lines connecting the data points are included as guides to the eye.}
\label{dlateralwater}
\end{figure}

\begin{figure}[t]
\includegraphics[width=\columnwidth]{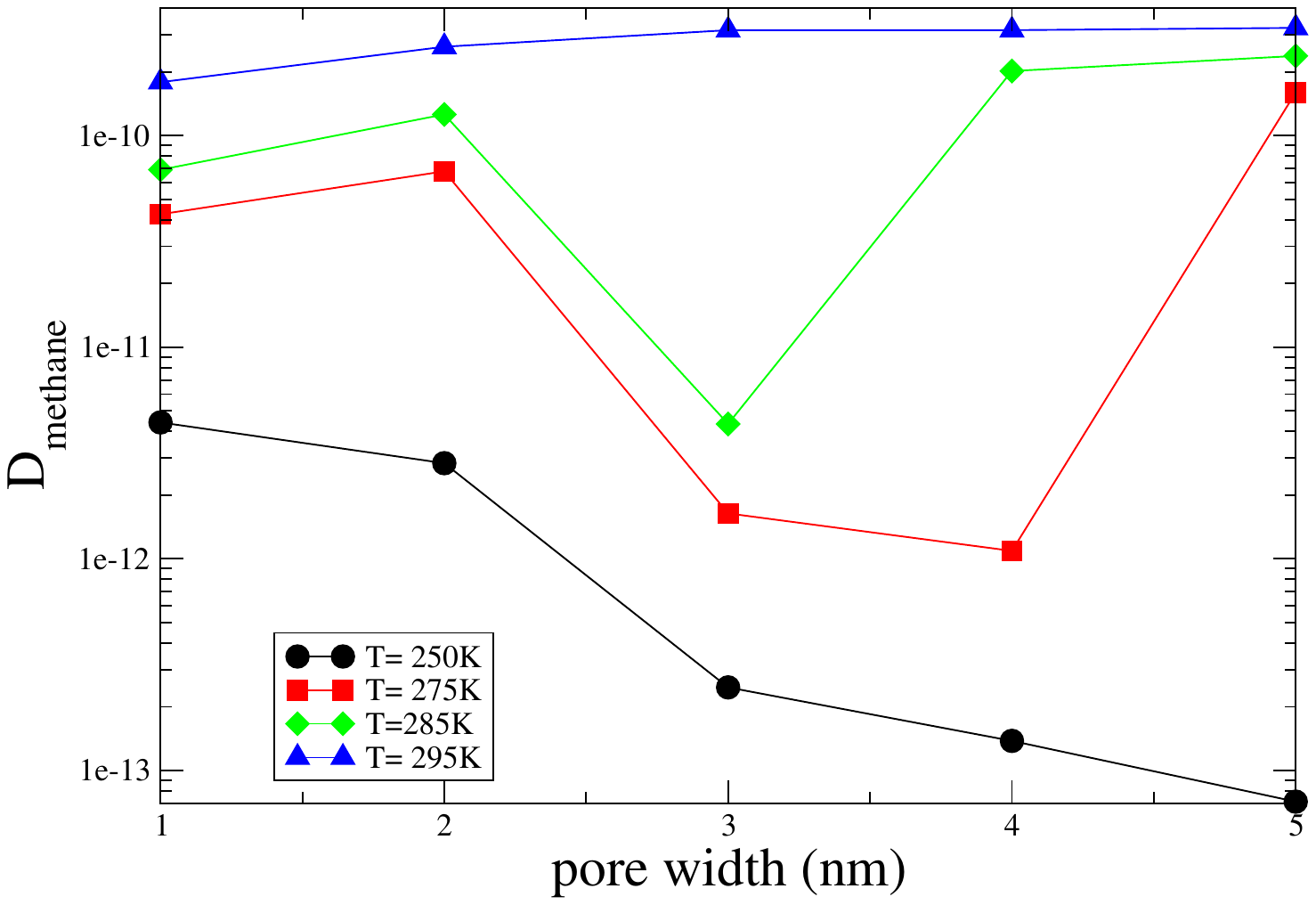}
\caption{Parallel diffusion coefficients of confined methane as a function of pore size for temperatures $T=250$, $275$, $285$, and $295$ K. The diffusion coefficients are shown on a logarithmic scale. From bottom to top, the datasets correspond to increasing temperature. Lines connecting the data points are included as guides to the eye.
}
\label{dlateralmethane}
\end{figure}

The parallel diffusion coefficients of water and methane exhibit similar dependencies on pore size and temperature (Figs.~\ref{dlateralwater} and \ref{dlateralmethane}), indicating that in-plane transport is primarily governed by the structural and dynamical state of the confined water matrix rather than species-specific effects. At $T=250$ K, both species show an increase in diffusion with decreasing pore size, consistent with confinement-induced restructuring of the hydrogen-bond network, where reduced three-dimensional connectivity and persistent in-plane correlations enhance lateral mobility. At intermediate temperatures ($T=275$ K and $T=285$ K), diffusion becomes non-monotonic, with a minimum at intermediate confinement, reflecting a regime of enhanced structural frustration and dynamical heterogeneity. At $T=295$ K, a monotonic decrease of diffusion with increasing confinement is recovered, consistent with conventional confined-liquid behavior dominated by wall interactions. Overall, methane mobility remains strongly coupled to the collective dynamics of the water network.
Unlike the in-plane direction, the mean squared displacement does not exhibit a clear asymptotic linear regime, preventing a consistent determination of diffusion coefficients $D_z$. Instead, the dynamics follow an anomalous scaling $\mathrm{MSD}(t) \propto t^{\alpha}$ with $\alpha < 1$, indicating subdiffusive behavior arising from confinement-induced trapping, spatial heterogeneity, and restricted connectivity. In this framework, the exponent $\alpha$ serves as a more appropriate descriptor than $D_z$.
The exponent $\alpha$ is obtained from the slope of $\log(\mathrm{MSD}(t))$ versus $\log(t)$, evaluated over the approximately linear region of the final 100 ns of the simulations. It should be interpreted as an effective quantity, $\alpha = \alpha(T, L, t_{\mathrm{fit}})$, reflecting the dependence of the observed scaling on temperature, confinement, and the selected time window.

\begin{figure}[t]
\includegraphics[width=\columnwidth]{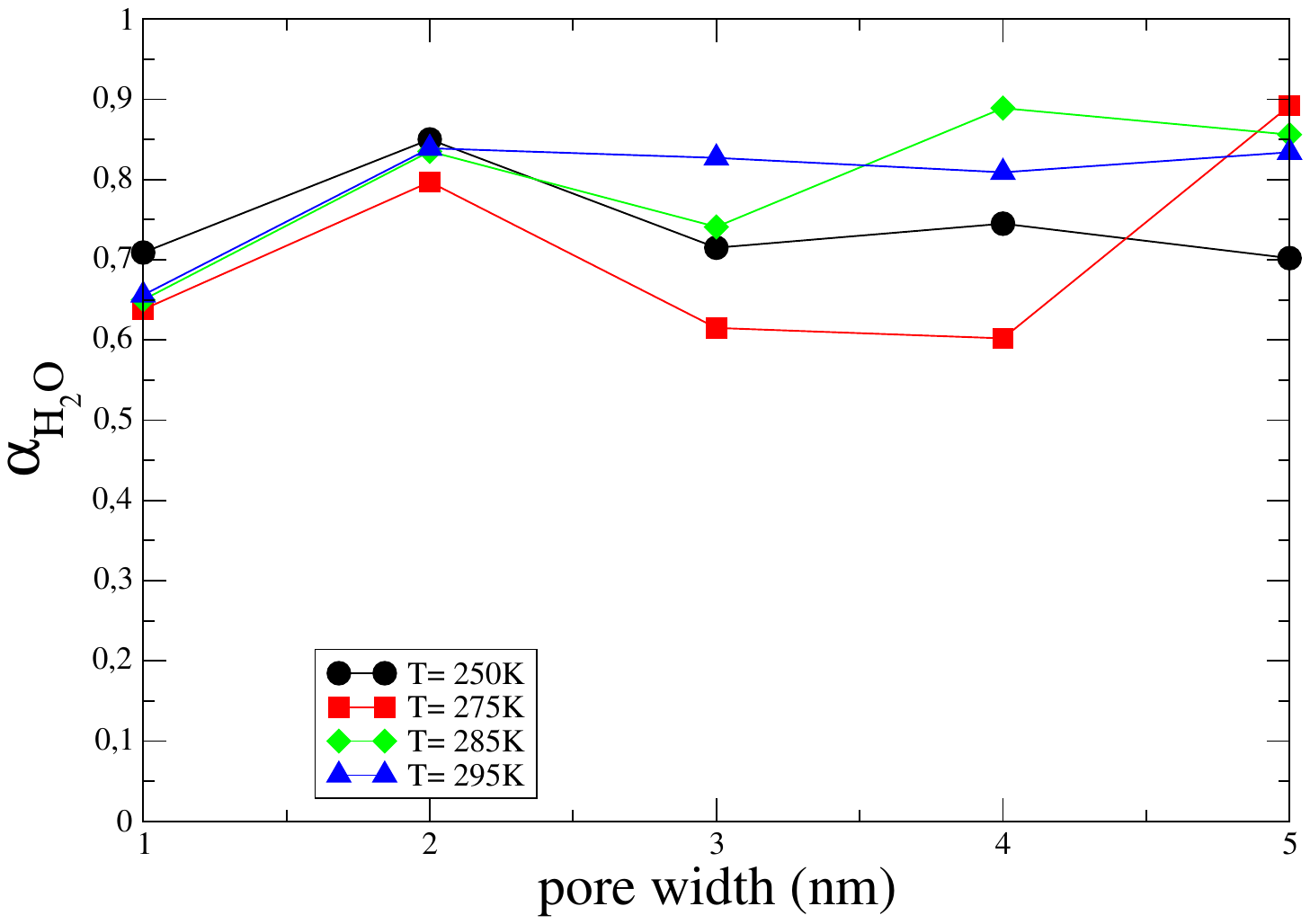}
\caption{Anomalous diffusion exponents, $\alpha$, associated with the perpendicular dynamics of confined water as a function of pore size for temperatures $T=250$, $275$, $285$, and $295$ K. The values of $\alpha$ remain below unity for all investigated conditions, indicating subdiffusive behavior in the direction normal to the confining walls. Lines connecting the data points are included as guides to the eye.}
\label{alphaw}
\end{figure}

\begin{figure}[t]
\includegraphics[width=\columnwidth]{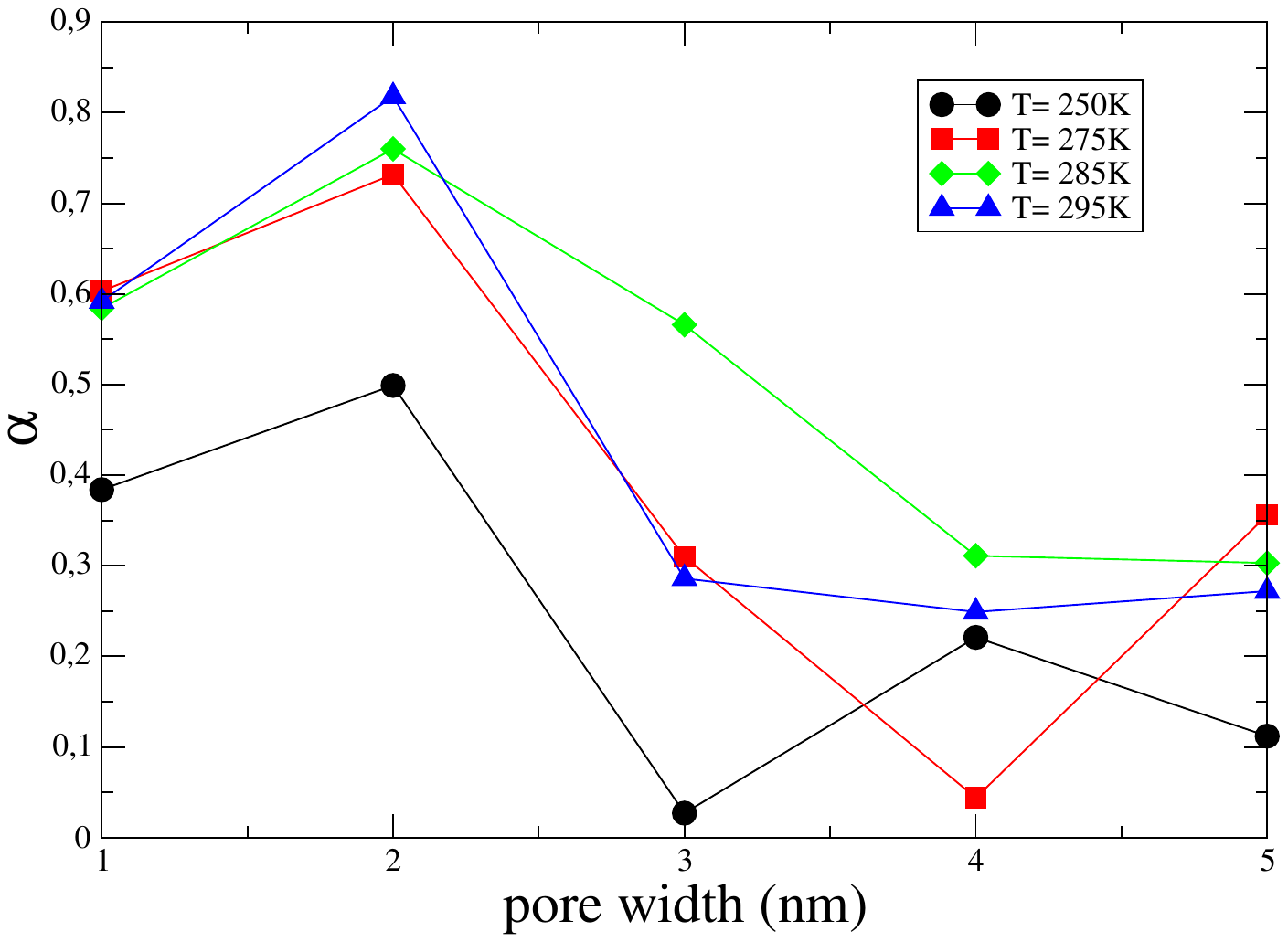}
\caption{Anomalous diffusion exponents, $\alpha$, associated with the perpendicular dynamics of confined methane as a function of pore size for temperatures $T=250$, $275$, $285$, and $295$ K. The values of $\alpha$ remain below unity for all investigated conditions, indicating subdiffusive behavior in the direction normal to the confining walls. Lines connecting the data points are included as guides to the eye.}
\label{alpham}
\end{figure}

\begin{figure}[t]
\includegraphics[width=\columnwidth]{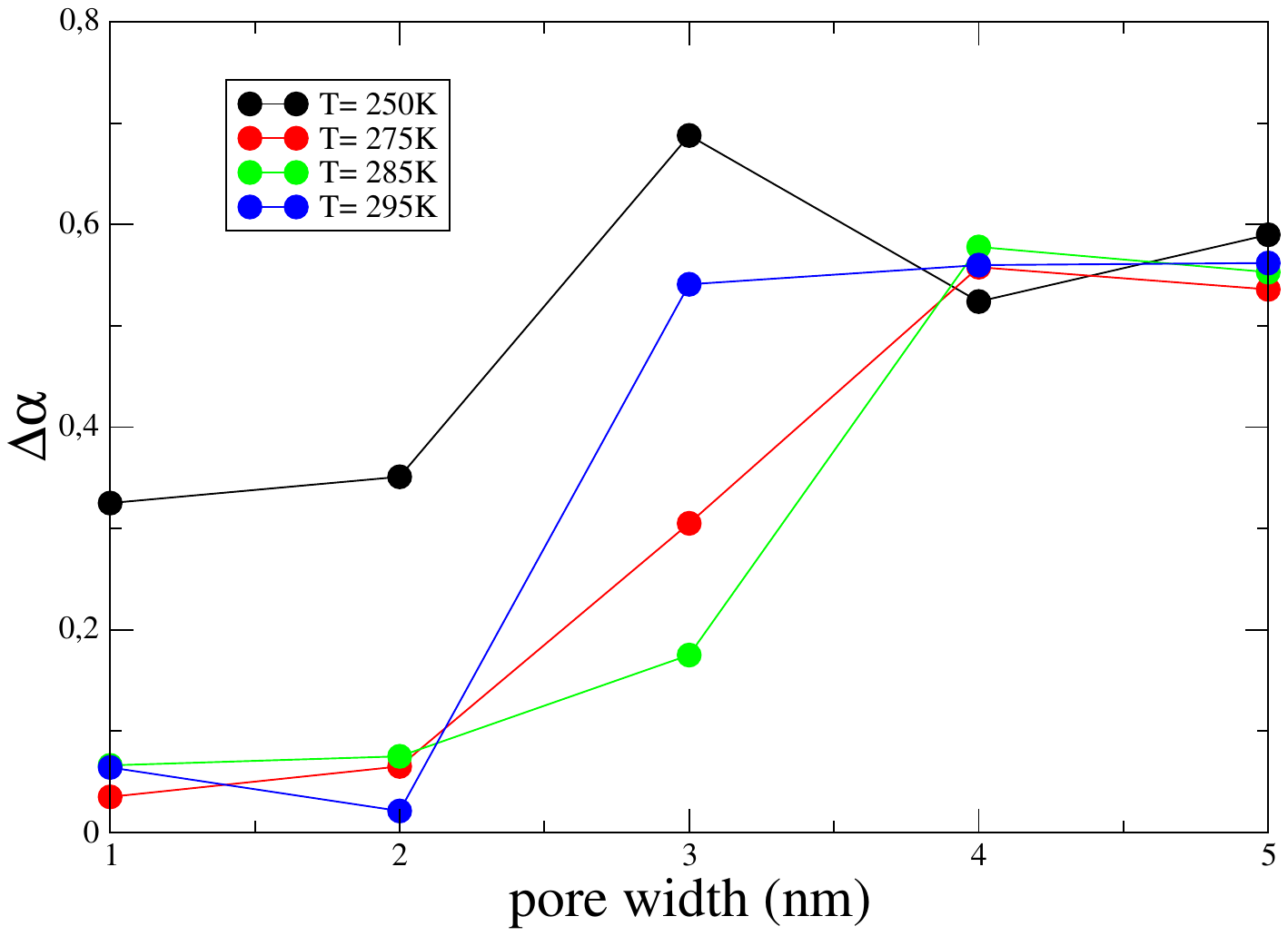}
\caption{Difference between the anomalous diffusion exponents, $\Delta \alpha = \alpha_{\mathrm{water}} - \alpha_{\mathrm{methane}}$, as a function of pore size for temperatures $T=250$, $275$, $285$, and $295$ K. This quantity is used as a metric to quantify the dynamical decoupling between confined water and methane in the direction perpendicular to the confining walls. Lines connecting the data points are included as guides to the eye.
}
\label{deltaalpha}
\end{figure}

Figure~\ref{alphaw} shows the anomalous diffusion exponent $\alpha$ for water as a function of pore size and temperature. In all cases, $\alpha < 1$, confirming subdiffusive transport perpendicular to the confining walls. In contrast to parallel diffusion, the temperature dependence of $\alpha$ is weak and non-systematic, with partial overlap between curves, indicating that perpendicular dynamics arise from the interplay of confinement-induced trapping, interfacial layering, and structural heterogeneity. For the smallest pore sizes (1–2 nm), $\alpha$ remains nearly constant across temperatures, suggesting that the dynamics are dominated by geometric constraints and reduced dimensionality. At intermediate pore sizes, $\alpha$ exhibits non-monotonic behavior, consistent with a regime of enhanced dynamical heterogeneity characterized by intermittent trapping. For the largest pore size (5 nm), $\alpha$ increases toward values closer to normal diffusion, although at 250 K it remains reduced, reflecting the persistence of structurally ordered environments that enhance trapping.
Figure~\ref{alpham} presents the corresponding results for methane. In all cases, methane exhibits systematically lower $\alpha$ values than water, indicating stronger subdiffusive behavior and a higher sensitivity to confinement-induced heterogeneity. This reflects its diffusion within a dynamically evolving water matrix rather than participation in the hydrogen-bond network. The strongest reduction of $\alpha$ occurs at intermediate pore sizes (3–4 nm), consistent with the regime of maximal structural frustration, where heterogeneous environments give rise to intermittent, trapping-dominated dynamics. A clear separation is observed for the 250 K data, indicating that enhanced structuring of water at low temperature significantly suppresses methane mobility. At larger pore sizes (4–5 nm), the curves for different temperatures converge, reflecting a partial recovery of a more homogeneous dynamical regime and a reduced influence of confinement-induced heterogeneity.

Figure~\ref{deltaalpha} reports the difference in anomalous diffusion exponents between water and methane, $\Delta \alpha = \alpha_{\mathrm{water}} - \alpha_{\mathrm{methane}}$, providing a direct measure of solvent–solute dynamical decoupling perpendicular to the confining walls.
In the smallest pores (1–2 nm), $\Delta \alpha$ remains relatively small and shows weak temperature dependence for $275$–$295$ K, indicating that both species experience similarly constrained dynamics under strong confinement. In this regime, geometric restrictions and reduced dimensionality dominate, affecting water and methane in a comparable manner. At $250$ K, $\Delta \alpha$ increases moderately, reflecting the enhanced impact of water structuring on methane mobility.
At intermediate pore size (3 nm), $\Delta \alpha$ exhibits a pronounced temperature dependence, reaching its maximum at $250$ K. This corresponds to the regime of strongest structural heterogeneity and dynamical frustration, where methane is more strongly affected by trapping and intermittent dynamics than water.
For larger pores (4–5 nm), $\Delta \alpha$ converges toward similar values across all temperatures, indicating a recovery of a more homogeneous dynamical regime. In this limit, confinement-induced heterogeneity is reduced and the dynamics of water and methane become increasingly coupled.
Overall, $\Delta \alpha$ shows a non-monotonic dependence on confinement, with a clear maximum at intermediate pore sizes where structural frustration and dynamical heterogeneity are most pronounced.

\section*{Conclusions}
In this work, we have investigated the structural organization and transport properties of methane–water systems under nanoconfinement, spanning pore sizes from strongly confined to quasi-bulk regimes. By combining density profiles, three-dimensional and in-plane radial distribution functions, local order parameters, and diffusion analysis, we have established a consistent picture in which confinement controls both structural ordering and dynamical behavior in a tightly coupled manner.

The density profiles reveal a gradual crossover from strongly interface-dominated configurations at small pore sizes to weakly modulated, quasi-bulk-like regimes at larger confinements. This structural evolution is mirrored in the local order parameters. While tetrahedral coordination ($F_3$) partially recovers as confinement is reduced, cubic-like ordering ($F_4$) exhibits a non-monotonic dependence on pore size and temperature, reaching its most pronounced expression at intermediate confinement. This indicates that structural order is not simply enhanced or suppressed by confinement, but instead optimized at specific length scales.

Radial distribution functions further clarify the dimensional nature of this ordering. Three-dimensional RDFs show the emergence of solid-like correlations under intermediate and weak confinement at low temperature, whereas these correlations are strongly suppressed at the smallest pore sizes. In contrast, projected RDFs demonstrate that lateral ordering persists even when three-dimensional correlations are absent, evidencing a confinement-induced reduction in the effective dimensionality of structural organization.

Transport properties exhibit closely related trends. Parallel diffusion coefficients display a non-monotonic dependence on pore size, reflecting the competition between confinement-induced mobility enhancement and structural frustration. Methane diffusion closely follows that of water, indicating that its dynamics are largely governed by the surrounding hydrogen-bond network. In the perpendicular direction, transport is consistently subdiffusive for both species, with anomalous exponents $\alpha < 1$ reflecting strong confinement-induced trapping and layering effects. Methane systematically exhibits lower $\alpha$ values than water, highlighting its greater sensitivity to the heterogeneous environment.

The dynamical coupling between both species, quantified through $\Delta \alpha = \alpha_{\mathrm{water}} - \alpha_{\mathrm{methane}}$, reveals a clear non-monotonic dependence on confinement. At the smallest pore sizes, both species are similarly constrained and dynamical decoupling remains weak. At intermediate confinement, $\Delta \alpha$ reaches a pronounced maximum, corresponding to the regime of strongest structural heterogeneity and frustration. At larger pore sizes, the dynamics progressively converge as confinement effects weaken.

Our results demonstrate that nanoconfinement induces a competition between confinement-driven local ordering and the frustration of extended three-dimensional structural correlations. This competition defines a characteristic confinement length scale at which structural order, dynamical heterogeneity, and solvent–solute decoupling are simultaneously maximized. Below this length scale, strong confinement suppresses three-dimensional correlations, leading to frustrated ordering despite the persistence of local structural motifs. These findings provide a unified microscopic framework for understanding how nanoconfinement governs the coupled structural and dynamical behavior of methane–water systems, with direct implications for the stability and formation of clathrate-like phases in confined environments.

\section*{Author contributions}
All authors contributed equally to this work. All authors have read and approved the final version of the manuscript.

\section*{Conflicts of interest}
There are no conflicts to declare.

\section*{Data availability}

The data supporting the findings of this study are available from the corresponding author upon reasonable request.

\section*{Acknowledgements}
MMP and AMFF acknowledge grant Refs.~PID2021-125081NB-I00 and PID2024-158030NB-I00 funded by MCIN/AEI/10.13039/501100011033 (Spain), FEDER EU and Programa de axudas de apoio á etapa de formación posdoutoral da Xunta de Galicia (Consellería de Cultura, Educación, Formación Profesional e Universidades) (orde 27/12/2023) P.P. 0000 421S 140.08. We also  acknowledge computing resources provided by CESGA (Santiago de Compostela, Spain). MPR acknowledges grant PID2023-151751NB-I00 funded by MCIN/AEI/10.13039/501100011033 and PID2024-158030NB-I00 funded by MCIN/AEI/10.13039/501100011033 (Spain). JTA would like to thank the Departamento de F\'isica Aplicada of the Universidade de Vigo for all the facilities provided  to a sabbatical stay, during which, this work was developed. Also to financial support of Universidad de Guanajuato and SECIHTI (through SNII), that makes this stay possible.

\bibliography{biblio} 
\bibliographystyle{rsc} 

\end{document}